\documentclass{aastex}          
\usepackage{spr-astr-addons}    

\usepackage[english]{babel}
\usepackage{amsmath}
\usepackage{amssymb}
\usepackage{epsf}
\usepackage{bm}
\usepackage{xcolor}
\usepackage{ulem}

\def\la{\; \raise0.3ex\hbox{$<$\kern-0.75em\raise-1.1ex\hbox{$\sim$}}\;}
\def\ga{\;  \raise0.3ex\hbox{$>$\kern-0.75em\raise-1.1ex\hbox{$\sim$}}\;}

\newcommand{\beq}{\begin{equation}}
\newcommand{\eeq}{\end{equation}}

\newcommand{\emaila}{ddofengeim@gmail.com, s.balashev@gmail.com, iav@astro.ioffe.ru}

\begin{document}
%
\title{Effect of a partial coverage of quasar broad-line regions
by intervening H$_2$-bearing clouds}

\shorttitle{Effect of a partial coverage}
\shortauthors{Ofengeim et al.}


\author{D.~D.~Ofengeim\altaffilmark{1,2}} 
\and
\author{S.~A.~Balashev\altaffilmark{2,3}} 
\and
\author{A.~V.~Ivanchik\altaffilmark{2,3}} 
\and
\author{A.~D.~Kaminker\altaffilmark{2,3}} 
\and
\author{V.~V.~Klimenko\altaffilmark{2,3}} 
\email{\emaila} 
%

\altaffiltext{1}{St.-Petersburg Academic University, 8/3 Khlopina street,
194021 St.-Petersburg, Russia}
\altaffiltext{2}{Ioffe Physical-Technical Institute,
Politekhnicheskaya 26, 194021 St.-Petersburg, Russia}
\altaffiltext{3}{St.-Petersburg State Polytechnical University,
Politekhnicheskaya 29, 195251 St.-Petersburg, Russia}


\begin{abstract}

We consider the effect of a partial coverage 
of quasar broad-line regions (QSO BLRs) by
intervening H$_2$-bearing clouds when a part of 
quasar (QSO) radiation passes by a cloud not taking part 
in formation of an absorption-line system  
in the QSO spectrum. 
That leads to modification of observable 
absorption line profiles and consequently to a bias 
in physical parameters derived from  
standard absorption line analysis.
In application to the H$_2$ {absorption} systems the effect 
has been revealed 
in the analysis of
H$_2$ absorption system in the spectrum of 
Q~1232+082 (see \citealt{Ivanchik2010}, 
\citealt{Balashev2011}). 
We estimate a probability of the effect 
to be detected in QSO spectra. 
To do this  we  derive  distribution of BLR sizes 
of high-z QSOs  from 
Sloan Digital Sky Survey (SDSS) Data Release 9 (DR9) 
catalogue and assume different distributions of cloud sizes. 
We
conclude that 
the low limit of the probability 
is about $11\%$.  The latest researches  
shows that about a 
fifth of observed 
H$_2$ absorption systems can be partially covered. 
Accounting of  the effect  may  allow  
to revise significantly
physical parameters of interstellar clouds  
obtained  by the spectral analysis.

\end{abstract}


\section{Introduction}
\label{introduct}

An effect of the partial coverage occurs
when an absorbing cloud 
does not fully cover an emission source. It is
mostly  probable at a situation in which
an angular size
of the source is comparable
with a size of the absorbing object.
The partial coverage is greatly important
for spectroscopic studies
including a choice of
an appropriate
model of the absorption-line systems (ALSs),
which is necessary to derive
physical parameters of an absorbing cloud
from  spectral  analysis. 
In the  simplest case of partial coverage the part
of  radiation from the emission source passes
by the absorbing cloud. It results in an additional
residual flux at the bottom
of absorption lines
registered in the spectrum 
with respect to the zero-flux level.
This effect is more pronounced
in the case of highly saturated
lines.

It should be emphasized that the most typical
situation for investigations of interstellar
medium is when angular size of an absorbing cloud
is much larger than an angular size of a background source
which can be considered as a point-like
object. In such cases
the effects of the partial coverage
are  very unlikely  and the observer
uses simple absorption-line
profiles defined by the optical depth
$\tau(\lambda)$
according to the
Bouguer-Lambert-Beer law.
However, this approach ceases
to be universal for studies
of the ALSs in  QSO  spectra, especially for relatively 
compact H$_2$ clouds 
whose
sizes can be comparable with sizes of QSO emission regions. 

QSOs are the
powerful active galactic nuclei 
located at cosmological distances (up to z$\sim$7).
Their high luminosity allows to detect
their spectra and to study medium 
between QSOs and the observer
via analysis of 
the ALSs imprinted in their spectra.
Remind that the ALSs detected in QSO spectra
can be classified in two general
categories: {\it intrinsic} and {\it intervening}.
{\it Intrinsic} absorption systems are related
to a region in the close vicinity of the central QSO
machine or 
the host galaxy.
{\it Intervening} absorption systems 
appear from
intergalactic medium and interstellar medium
of galaxies 
remote 
from the QSO. 
It was found that the effects of partial
coverage, breaking the universal conception
of a point-like source against
an extended absorber,
is a rather frequent phenomenon 
for the intrinsic ALSs
in QSO spectra \citep{Petitjean1994}.

However, up to recently in the case
of intervening ALSs the effects of partial
coverage were believed to be unlikely.
Actually, the projected sizes of typical
intervening absorbers
are much larger
than quasar emission regions. 
The emission regions of QSOs 
in UV and optical range
have been  constrained
by a size of $\la$ 1~pc.
For example, reverberation mapping establishes the relationship 
between the BLR size and the luminosity and yields a BLR size 
of ${\rm R}_{\rm BLR}\sim$\,0.2\,pc \citep{Kaspi2007, Chelouche2012} 
for high redshift luminous quasars. Differential microlensing 
allows one to estimate the size of the BLR $\sim 0.1$ pc \citep{Sluse2011}. 
The observations of gamma-ray emission constrain the size of the jet 
extension to a few parsecs \citep{Abdo2010}.
On the other hand,  majority of ALSs 
constitute the Ly-$\alpha$ forest, which
are related
to intergalactic
clouds with sizes well
exceeding 10 kpc \citep{Rauch1998}.
Sizes of
diffuse atomic clouds inside
intervening galaxies 
are considered
to imprint 
the damped
Ly-$\alpha$ systems (DLAs), 
or
sub-DLAs,
can be roughly estimated 
as $10$~pc -- $1$~kpc.
At last, 
the
Lyman-limit systems (LLS), 
as well as
Mg~II and C~IV  absorbers,
seems to correspond also
to kpc-scale clouds in the halo
of the intervening galaxies.

Nevertheless, the partial coverage of Mg~II
system in the spectrum of the QSO, APM~08279+5255,
was firstly reported by \citet{Ledoux1998}.
However,  it had been found some
later that this was
very peculiar case of the partial coverage
due to a gravitational lensing
of the background QSO 
(see \citealt{Ledoux1998,Ibata1999}).
More definitely the effect of
partial coverage
has been revealed
in the case of H~I
21cm absorption lines measurements
in the DLA system  \citep{Kanekar2009}.
Actually, the emission regions
of radio-loud QSOs
at a wavelength 21 cm
can well exceed a size of 10 kpc
and typically consist
of compact few pc cores and extended
dilute envelopes.
The effect of partial
coverage essentially alters derived spin
temperature T$_{s}$  of the neutral
gas absorber \citep{Kanekar2009}.

Molecular bearing ALSs
are another type of currently observed absorption
systems in QSO spectra
which may have projected sizes
comparable with a size of QSO emission region.
These systems include molecular
absorption lines (predominantly
H$_2$, in some cases --
HD and CO)
and refer to the diffuse
and translucent molecular phases
of the interstellar medium \citep{Snow2006}.
Estimated number density
of this system is {$\gtrsim$} 10 cm$^{-3}$ at the
column density $\la 10^{20}$~cm$^{-2}$,
consequently these clouds constrained by pc or
even sub-pc sizes (see Sect.\ref{Distr-Geom}).

The major part of QSO radiation
is generated by 
an accretion disk (AD)
around the black hole
and forms a continuous spectrum.
Characteristic size of 
the AD is considered
to be $\la 3\times10^{-3}$~pc 
(e.g. \citealt{Blackburne2011, JimenezVicente2012}). 
Thus this source may be treated as a point-like
object fully covered by molecular clouds. 
On the other hand, there is still an extended
region around the AD of relatively dense and warm
partly ionized gas,
called 
a broad-line region (BLR), where
broad emission lines  are formed, that leads to  
some additional 
enhancement of the continuum in QSO spectra.
The angular
size of this region at least an order
of magnitude exceeds the typical angular 
sizes of 
ADs and may be
comparable with sizes of H$_2$ molecular clouds.
Therefore an incomplete coverage of BLRs 
by molecular clouds situated on the line-of-sight
between the QSO and observer becomes
quite probable.

The partial coverage of a BLR
by the intervening
H$_2$ absorption cloud was reported
by
\citet{Ivanchik2010} and investigated in details
by \citet{Balashev2011}. It was demonstrated that the
H$_2$-bearing cloud ($z_{abs}=2.34$)
covers completely the QSO\, 1232+082
($z_{em} = 2.57$)
intrinsic continuum source,
but only a part of the BLR.  

There are also other possibilities 
for the partial coverage effect: (i) jet emission can be 
scattered/re-emitted by  medium  in the quasar host galaxy, 
which is remote from the AD and BLR. 
In this case 
the observer 
detects the partial coverage of a continuous spectrum. 
This is the most probable explanation of the
recently revealed 
partial coverage of the  Q\,0528-250 
continuous spectrum 
by H$_2$ absorption system at z=2.811 \citep{Klimenko2015}. 
(ii) The calculation 
of galaxies evolution shows that a galaxy at high redshifts 
can have several supermassive black holes (SMBHs;\,  
binary, triple or quadruple systems of SMBHs, \,
e.g.\,\citealt{Kulkarni2012}), 
in particular, one of SMBHs can remain unscreened by an absorbing molecular cloud. 
(iii) 
The observer can register UV emission 
of 
a QSO host galaxy much more extended than a BLR.  
(iv) 
The partial coverage can be caused 
by gravitational lensing 
(e.g. the case of APM\,08279$+$5255)
of a galaxy or a galaxy cluster 
located between 
a quasar and intervening clouds.

The effect of 
partial coverage leads to serious 
changes of parameters  
derived from an analysis  of QSO spectra 
(e.g. column densities of absorbers, 
\citealt{Ivanchik2010, Balashev2011, Klimenko2015}).
So  it is worthwhile
to estimate 
a probability 
to find 
the partial coverage of a QSO BLR 
by an intervening H$_2$ absorption system. 
Other 
appearances
of the partial coverage (listed above), 
apparently, lead 
only to increase the probability.

Following \citet{Balashev2011}, one can introduce a 
{\it flux coverage factor} $f$ 
as the ratio of a light flux 
comprising the absorption systems imprinted
in its spectrum
to the total flux from 
a quasar. 
In the present paper 
we calculate a probability to reveal 
the partial coverage of a QSO BLR 
located at some redshift $z_q$ by  H$_2$ ALS 
at redshift $z_c$ (cloud) with coverage factor $f < f_0$ 
($f_0$ is  a fixed value, see below).
We consider an arbitrary angular distance 
between  line-of-sights 
to the centres of the cloud 
and  BLR, as well as an arbitrary 
ratio $\kappa$ of the cloud to BLR 
transverse sizes. 

The paper significantly develops our previous 
work \citep{Ofengeim2013} and  is organized as following. 
At first, we 
proceed from an
oversimplified assumption
of uniformity
of the radiation flux from
the whole BLR region
ignoring
the luminous point-like AD in the centre
of the QSO. 
In Section~\ref{Geom} we introduce
a geometrical coverage
factor $g \equiv f$ 
and consider the main parameters
controlling $g$. 
In Section~\ref{Distr-Geom}
we calculate
a probability to reveal factor $g < g_0$,
where $g_0$ is also a fixed value 
(an analogue of $f_0$),
for certain distributions 
of the main parameters.
In Section~\ref{Distr-Flux}
we discuss similar 
probability but for more realistic
coverage factor $f$ determined 
as the
ratio of radiation fluxes, 
taking into account
nonuniformity of the radiation flux and 
including
the intense AD radiation.
Sections~\ref{redshift}  and  \ref{part_cov} 
present  a special consideration 
of the dependence 
of the factor $f$ 
on the redshifts $z_q$ and $z_c$.
In Section~\ref{concl} we discuss  obtained
results and their relation to observations.

\section{Geometry of partial coverage}
\label{Geom}
\begin{figure*}[t]  
\centering
\includegraphics[width=\textwidth]{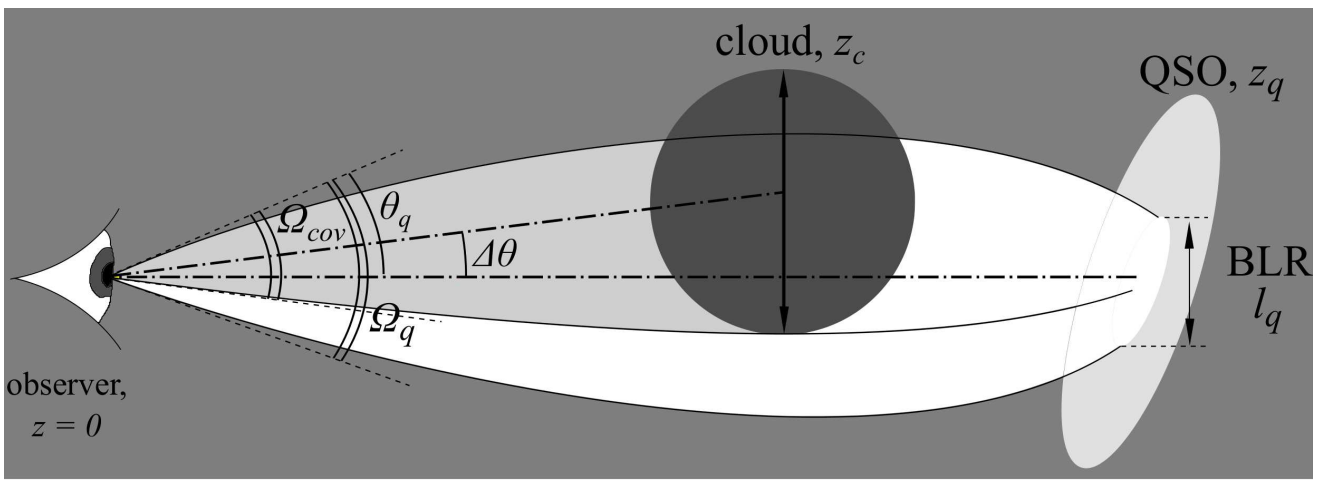}
\caption 
{Schematic illustration of an absorption
cloud at a transverse size $l_c$ (dark grey circle) 
situated between the observer and 
QSO BLR with a transverse size 
$l_q$ (white ellipse in the centre of 
QSO);  the wider ellipse
around BLR (light grey)  symbolizes a QSO host 
galaxy; $\theta_q$ is an angular size of BLR,
$\Delta\theta$ is an angle between two line-of-sights 
from the observer to the centres of 
QSO and cloud; $z_q$ and $z_c$ are 
redshifts  of the 
QSO and cloud; $\Omega_q$ is a solid
angle (light cone) of the whole BLR radiation 
flux, $\Omega_{cov}$ is a solid angle of a part 
of BLR flux passed through the cloud. 
Radiation from 
the QSO without traces of H$_2$-absorption
systems in its spectrum is shown 
as a white area of the light 
cone and  radiation containing the absorption
systems -- as a light grey area. In general 
case the light cone is curved due to the
expansion of the Universe. } 
\label{objects}
\end{figure*}

\begin{figure*}
\centering
\includegraphics[width=\textwidth]{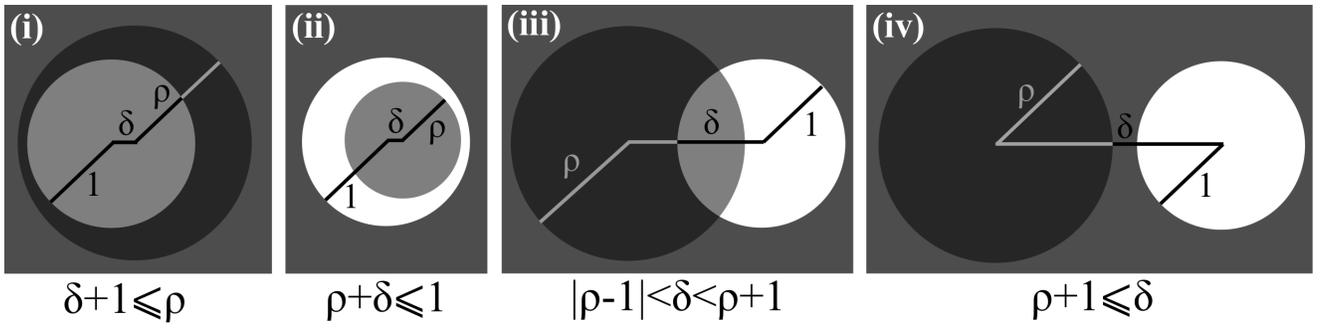}
\caption{
Four types of coverage (grey regions)
of QSO BLRs (white circles) 
by clouds (black circles) 
in the observer reference frame:
(i)  full coverage, $g=1$ (widely accepted case,\,
white circle coincides with grey one);\,
(ii) partial {\it ring-shaped} coverage, $0 < g < 1$
(black circle coincides with  grey one);\,
(iii) {\it crescent-like} coverage, $0 < g < 1$;\, 
(iv) full {\it noncoverage}, $g=0$.
All angular sizes are represented
in units of $\theta_q$:
the angular size (radius) of the BLR
is  1, the radius of the cloud is $\rho$, and the
angular deviation is $\delta$ (see text).
}
\label{types}
\end{figure*}

In this work we use a toy 
model to describe 
the emission regions of  QSOs and  absorbing clouds. 
The geometry of this model is presented 
in Fig.~\ref{objects},\ref{types}.
The transverse sizes
of the BLR and the cloud are designated
as $l_q$ and $l_c$,
the angular size of the BLR
is designated by $\theta_q$  
(hereinafter angles are in radians), 
a corresponding angle
for the cloud can be denoted as $\theta_c$
(is not shown in Fig.~\ref{objects}).
The light from the BLR propagates
through the Universe and it is registered
by the observer. The observer detects 
an emission within
a light cone restricted by 
an angle $\theta_q$ or a solid angle
$\Omega_q$ comprising
the whole flux from the BLR.
Some part of light from the 
BLR is partly shielded by the cloud
in such a way that 
molecules inside the cloud
imprint a set of absorption lines
(or ALS) in the initial spectrum.
This part is bounded by a solid
angle (of coverage) 
$\Omega_{cov}$. The rest
light can pass by the cloud
and come to the observer without
formation of ALSs in its spectrum. This flux of
radiation may be approximately estimated
as a flux comprised by a solid angle
$\Omega_{uncov} = \Omega_q - \Omega_{cov}$.
In result the observer detects
a complex radiation from the QSO with
spectrum integrated  over angles.

Let us determine a {\it geometrical}
coverage factor
$g$ as a simplest way to estimate factor $f$.
If we assume that 
fluxes from covered and uncovered parts of the BLR
are uniformly distributed within their
solid angles, then factor $g$ may be estimated
as a ratio of the solid angle 
$\Omega_{cov}$ to the whole solid
angle $\Omega_q$ :
\begin{equation}
g =  \frac{\Omega_{cov}}{\Omega_q},
\label{g}
\end{equation}
where at $\theta_q \ll 1$  we can write
$\Omega_q=\pi \theta_q^2$.
We imply that $g \equiv f$ in the case 
of pure geometrical estimations.
Actually, the factor $g$ may essentially
differ from $f$ because
it does not take into account nonuniformity
of fluxes within the solid angles
(see Section~\ref{Distr-Flux}).

Thereafter it is convenient to express
all angular values
in units of $\theta_q$ and all solid angles --
in units of $\Omega_q$. Thus a 
relative  angle between
directions to the  
centres of  QSO and cloud (angular deviation)
is determined as
\begin{equation}
\delta = {\Delta \theta \over \theta_q},
\label{delta}
\end{equation}
and the relative angle size of a cloud becomes
\begin{equation}
\rho = {\theta_c \over \theta_q}.
\label{rho}
\end{equation}

An angle size $\theta$ of an object
with a {transverse} size $l$ at the cosmological
redshift $z$ may be determined  by a standard
way (e.g. \citealt{Zeldovich1983}; \citealt*{Kayser1997}) as
\begin{equation}
\theta = {l \over {2\  D_A(z)}},
\label{theta}
\end{equation}
where $D_A(z)$ is 
an angular diameter distance.
In the standard $\Lambda$CDM
cosmological model the value
$D_A(z)$ 
to a cosmologically distant object
at the redshift $z$  is
(e.g. \citealt{Zeldovich1983, Kayser1997})
\begin{equation}
D_A(z) =
{c \over H_0} {1 \over 1+z} \int_0^z
{{\rm d}z' \over
\sqrt{\Omega_m(1+z')^3 +
\Omega_{\Lambda}}},
\label{DA}
\end{equation}
where
$c$ is the speed of light,
$H_0=100~h$~km~s$^{-1}$
is the present Hubble constant,
$c/H_0\ \approx\ 2.998~h^{-1}$~Gpc;
$\Omega_m$ is the dimensionless
matter density parameter and
$\Omega_{\Lambda}=\Lambda c^2/(8 \pi G)$
is determined by the cosmological constant
$\Lambda$, $G$ is the gravitational constant;
hereafter we assume that
$\Omega_m=0.3$ and
$\Omega_\Lambda=0.7$ (\citealt{Planck2013XVI})

Using Eq.~(\ref{theta}) one
can rewrite Eq.~(\ref{rho}) as
\begin{equation}
\rho(\kappa,\, z_q, z_c) = \kappa \, {D_A(z_q) \over D_A(z_c)},
\label{rho_kappa}
\end{equation}
where the relative angular size of the cloud
is expressed through
the ratio of transverse sizes of the cloud and QSO
\begin{equation}
\kappa = {l_c \over l_q}.
\label{kappa}
\end{equation}

The coverage factor $g$ is a function 
of $\rho=\rho(\kappa;\,  z_q, z_c)$ and $\delta$.
Fig.\,\ref{types} demonstrates
four basic types of
relative positions
of two circles conventionally
referred to the BLR and cloud
projected on the sky.

These types corresponding to
special dependence $g=g(\rho, \delta)$
can be formalized as follows:

\noindent
(i) {\it Full coverage}: $1+\delta \leq \rho$.
All flux from the QSO BLR
goes through the cloud, $g=1$. 
Up to recently, only this case has been considered
in literature in application to intervening H$_2$ clouds.

\noindent
(ii) {\it Partial ring-shaped coverage}: $\rho+\delta \leq 1$.
The angular size of the BLR 
exceeds the angular size of the
cloud plus the angular deviation. In this case $g= \rho^2$.

\noindent
(iii) {\it Crescent-like coverage}: $| \rho-1 | < \delta < \rho+1$.
After some calculations one can obtain
\begin{equation}
\begin{split}
g(\rho,\delta) &\equiv s(\rho,\delta) = \frac{\rho^2}{\pi}
\arccos {{\rho^2+\delta^2-1} \over 2 \rho \delta} \\
	&+ \frac{1}{\pi}\arccos{{1+\delta^2-\rho^2} \over 2 \delta} \\
	&- {1 \over 2\pi} \sqrt{[\rho^2-(\delta -1)^2][(\delta+1)^2-\rho^2]}.
\end{split}
\label{s_r_d} 
\end{equation}

\noindent
(iv) {\it Full noncoverage}: $1+\rho \leq \delta$.
Relative angular  distance is large and the cloud remains
unobservable in the QSO spectrum. In this case $g=0$.

One can summarize
all cases (i)--(iv):
\begin{equation}
g(\kappa, \delta;\, z_q, z_c) = \left\{
      \begin{array}{ll}
         1; &  {\rm (i)}    \\
    \rho^2(\kappa;\, z_q, z_c);  & {\rm (ii)}  \\
    s[\rho(\kappa;\, z_q, z_c), \delta]; & {\rm (iii)} \\
     0;  & {\rm (iv)}
\end{array}   \right.
\label{g_k_d}
\end{equation}

\section{Distribution of coverage factor}
\subsection{Geometrical coverage factor}
\label{Distr-Geom}

Let us calculate the cumulative distribution function 
(CDF) 
$P(g < g_0)$,
which describes the probability to detect an absorption
system in QSO spectrum with geometrical coverage 
factor $g$ not exceeding a value $g_0$.
Since one can reveal
the cloud  only detecting absorption lines in QSO spectra,
the case of full noncoverage $g=0$ is equivalent to the
absence of an observed cloud. Therefore we can put
(rather conventionally)  that
the probability of full noncoverage is $P(g=0)=0$.

More specifically, to detect an absorption system 
in a spectrum it is necessary that a cloud totally 
cover the QSO AD\footnote{Remind that the AD generates 
the most part of continuum radiation}.
In contrast, if the cloud partially cover only BLR and does not cover AD, 
H$_2$ molecular absorption lines {associated with the cloud} in the spectrum 
will be detectable only in surroundings of the top
of quasar emission lines 
blueshifted of Ly$\alpha$
(e.g. C\,{\sc iii} or Ly$\beta$+O\,{\sc vi}, 
see Fig.~\ref{spec}). However,
in such a case the effect of the partial covering
would be difficult
to identify, since only several weak absorption lines 
of the H$_2$ ALS could be expected. 
These lines would be hardly detectable against
the numerous Ly-alpha forest lines in the spectrum. 
It gives an additional restriction
on the angular deviation:
$\delta < \rho$, or $P(\delta > \rho)=0$.

Let us add, that the effect of the partial coverage
can be observed also for C\,{\sc iv} 
emission line (redshifted of Ly$\alpha$) 
using  C\,{\sc i} (neutral carbon) absorption lines 
associated with H$_2$  \citep{Ivanchik2010,Balashev2011}, 
but this case is not considered in this paper.

The  CDF $P(g < g_0)$ can be determined 
through the distributions 
of two random values $\kappa'$ and $\delta'$.
Let us determine the probability 
density functions (PDFs) for
these two distributions:  $p_{\kappa}(\kappa)$ and
$p_{\delta}(\delta|\,  \kappa)$, respectively,
the latter value being the conditional PDF of the
occurrence of $\delta'=\delta$ at a given value 
$\kappa'=\kappa$.
Following to the standard definitions of the
Probability theory
one can introduce the relations between these PDFs and
differential probabilities:
$P(\kappa < \kappa' < \kappa + {\rm d}\kappa)=
p_{\kappa}(\kappa)\  {\rm d}\kappa$  and
$P(\delta < \delta' < \delta + {\rm d}\delta)=
p_{\delta}(\delta|\,  \kappa)\ {\rm d}\delta$,
and as a consequence to write
\begin{multline}
P(g<g_0) = \\
= \int_0^\infty \int_0^\infty                      
             p_{\delta}(\delta|\,  \kappa)\,  p_{\kappa}(\kappa)\,
	     \Theta[g_0-g(\kappa,\delta)]\,
         {\rm d}\delta\,  {\rm d}\kappa. 
\label{P(g)}
\end{multline}
where $\Theta(x)$ is the Heaviside function. 
One can optimize calculations using the reciprocal relation 
$\kappa(g,\,\delta)$ given in 
Eqs. (\ref{k_a}) and (\ref{k_b}) of Appendix. Then integration over 
$\kappa $ in Eq.~(\ref{P(g)}) 
should goes from $0$ to $\kappa(g_0,\,\delta )$.

For simplicity, we assume the uniform mutual distribution
of QSO and cloud centres
on the sky and get $p_{\delta}(\delta) \propto \delta$.
To obtain a normalization function for $p_{\delta}(\delta)$
we use the condition $\delta < \rho$ 
of the reliable
detection of H$_2$  ALS  (see above). 
In result we obtain
\begin{equation}
p_{\delta}(\delta|\, \kappa) = 
                       {2 \delta \over [\rho(\kappa)]^2}\,
                       \Theta[\rho(\kappa)- \delta].
\label{p_delta}
\end{equation}
It was chosen that $z_q=2.57$, $z_c=2.33$ in all
calculations employing Eq.~(\ref{rho_kappa})
for $\rho(\kappa, z_q, z_c)$, 
i.e. they refer to the 
case of the partial coverage of 
Q\,1232+082 (\citealt{Balashev2011}). 
The dependence of 
results on
the redshifts 
will be investigated in Section~\ref{redshift}.

The PDF $p_{\kappa}$ with respect to $\kappa$ may be
represented as
\begin{equation}
p_{\kappa}(\kappa)= 
                  2 \int_0^\infty \, p_{cloud}(2 \kappa r)\,
                  \, p_{qso}(r)\ r{\rm d}r,
\label{p_kappa}
\end{equation}
where $p_{cloud}$ and $p_{qso}$ are PDFs for linear sizes of
clouds $l_c = 2 \kappa r$  and  radii $r=l_q/2$ of
BLRs, respectively; in derivation of Eq.~(\ref{p_kappa})
it is used that d$l_c=2 r {\rm d}\kappa$; factors d$\kappa$
are reduced in both sides of Eq.~(\ref{p_kappa}). 

To obtain $p_{qso}(r)$ we exploit an empirical equation
by \citet{Kaspi2005} between 
the radius $r$ and QSO luminosity
\begin{equation}
{r \over 10\, \text{light-days}} = A L^B,
\label{r_L}
\end{equation}
where $L=\lambda\, L_{\lambda}/ (10^{44}~[{\rm erg}/{\rm s}])$ 
is 
the normalized
QSO luminosity
at a fixed QSO-restframe wavelength 
$\lambda$, $A$ and $B$ - 
numerical coefficients.
To estimate QSO luminosity distribution 
we used  QSO spectra from  SDSS DR~9,\, (see \citealt{Ahn2012}). 
It is commonly accepted that due to high column density 
H$_2$-cloud associated with large amount of neutral hydrogen, 
i.e. associated with DLAs. 
We used the sample of DLAs bearing QSOs 
\citep{Noterdaeme2012} from 
the whole SDSS DR9 QSOs sample. 
Choosing in Eq.~(\ref{r_L}) 
luminosities 
at the QSO-restframe 
wavelength $\lambda =1450$\,\AA\ 
and  quantities $A=2.12$,
$B=0.496$ (errors omitted)
we can calculate a required distribution $p_{qso}(r)$.
\begin{figure*}
\centering
\includegraphics[width=0.495\textwidth]{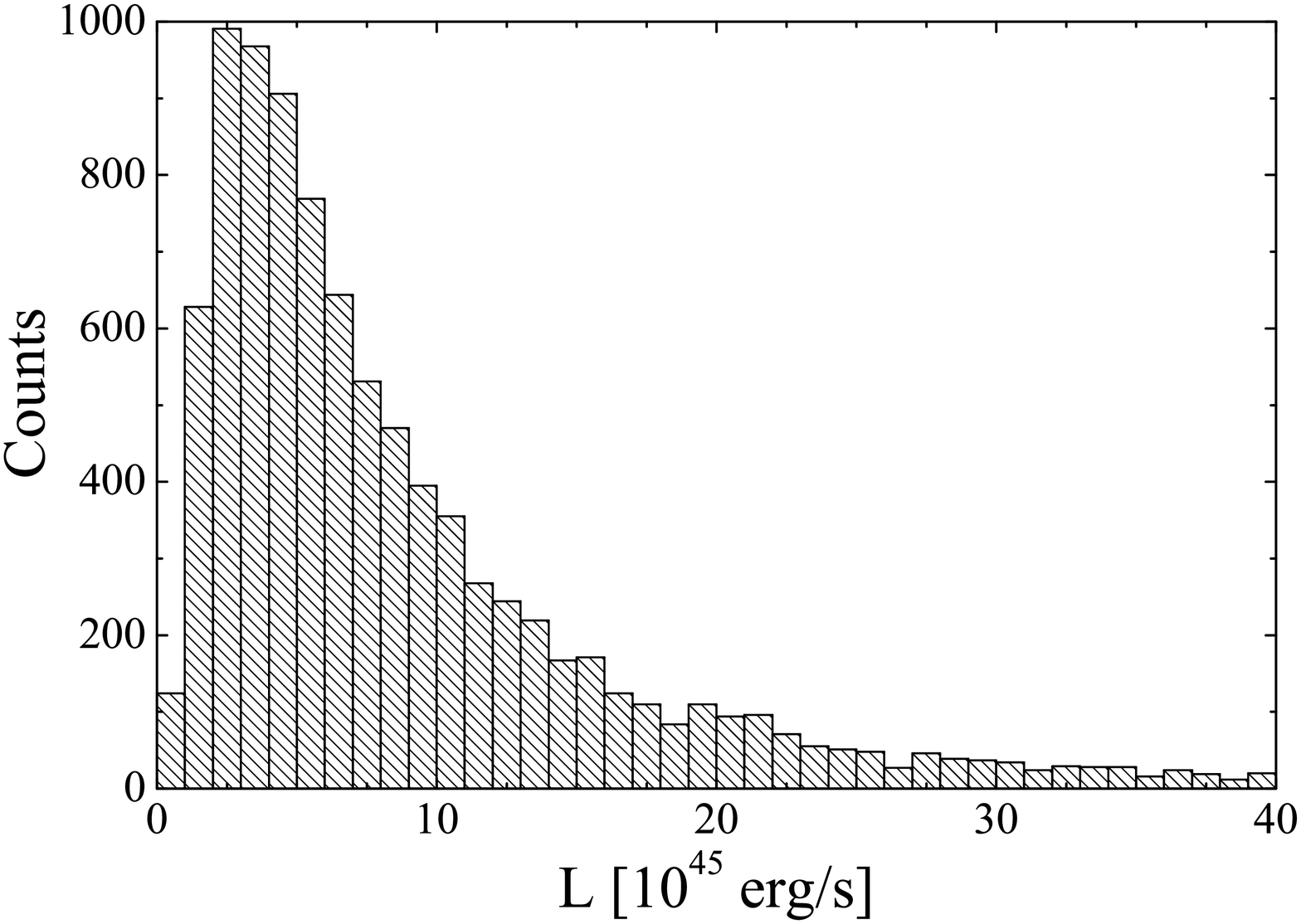}
\hfill
\includegraphics[width=0.495\textwidth]{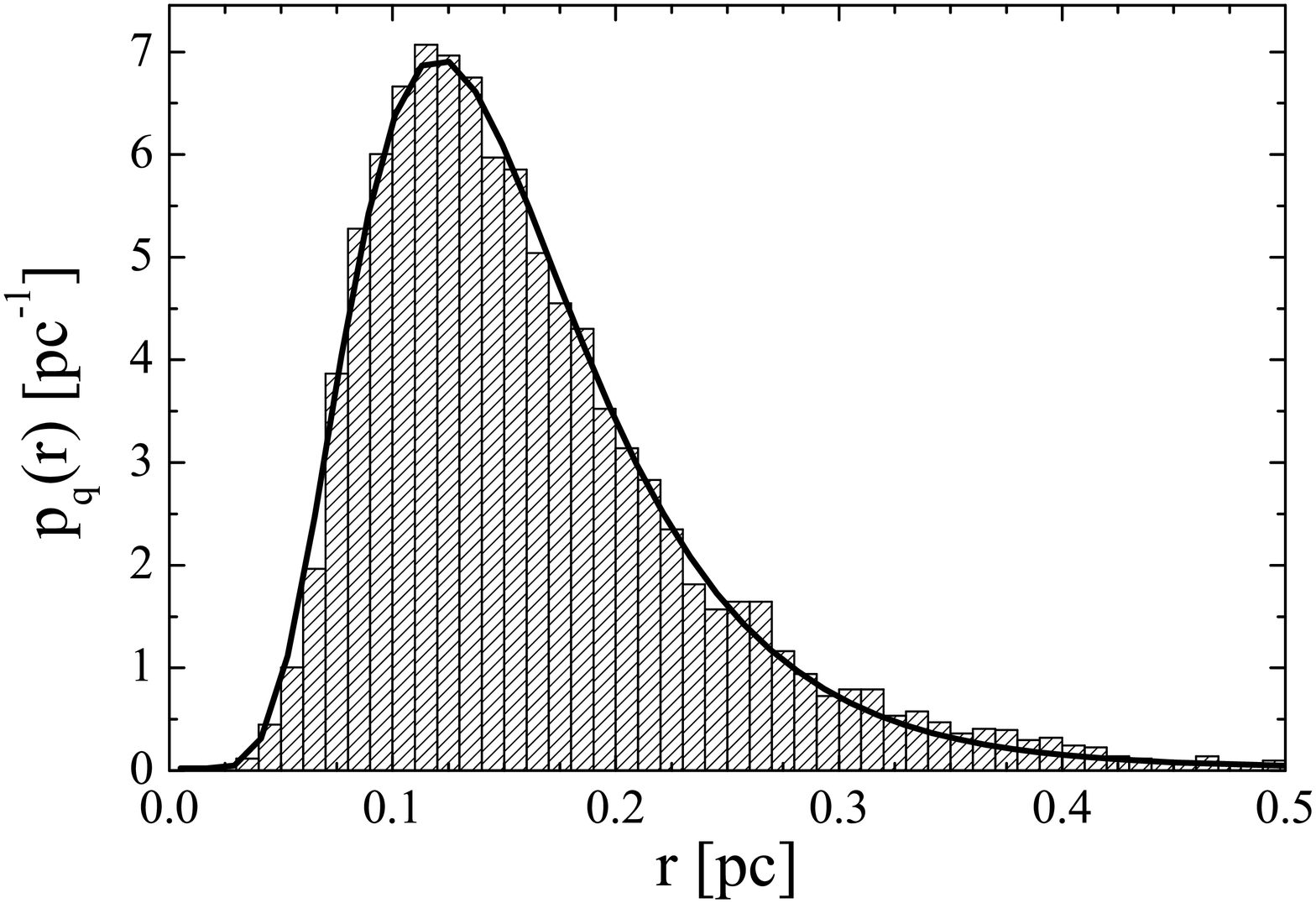}
\caption{
{\it Left panel}: Histogram (hatched bars)
of QSO-luminosities
sampled at a wavelength of ${\rm 1450\,\AA}$
from the data of SDSS\,DR9,
QSOs are chosen under condition of  the presence of DLAs 
in their spectra; {\it right panel}: 
probability density function (PDF)
of BLR radii, hatched histogram bars 
form a distribution obtained  with using
Eq.~\protect{(\ref{r_L})} and the data shown in the left panel;
solid line is the resulting
lognormal approximation.
}
\label{LumBLR}
\end{figure*}
%
\begin{figure*}
\centering
\includegraphics[width = 0.495\textwidth]{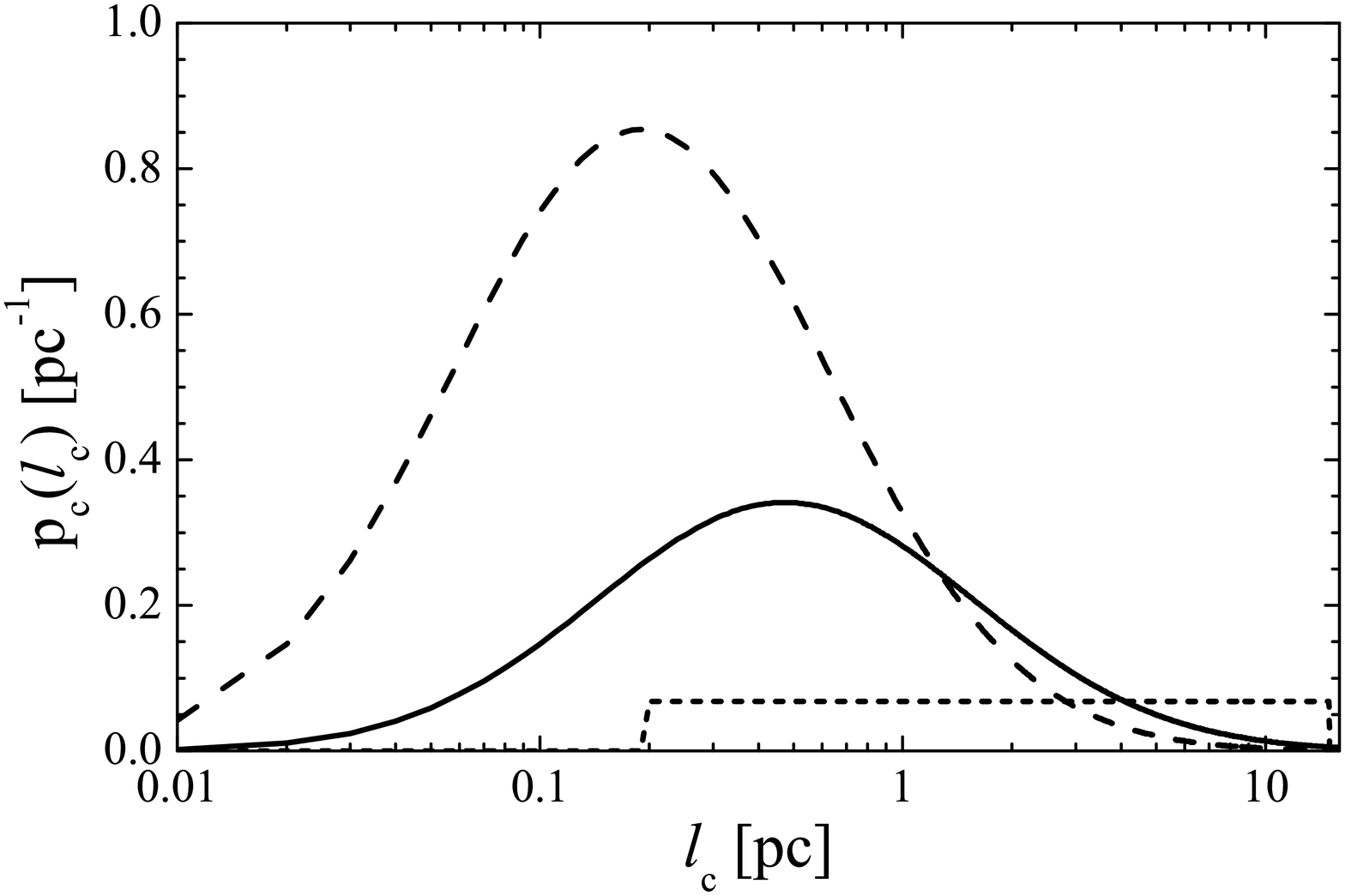}
\hfill
\includegraphics[width = 0.495\textwidth]{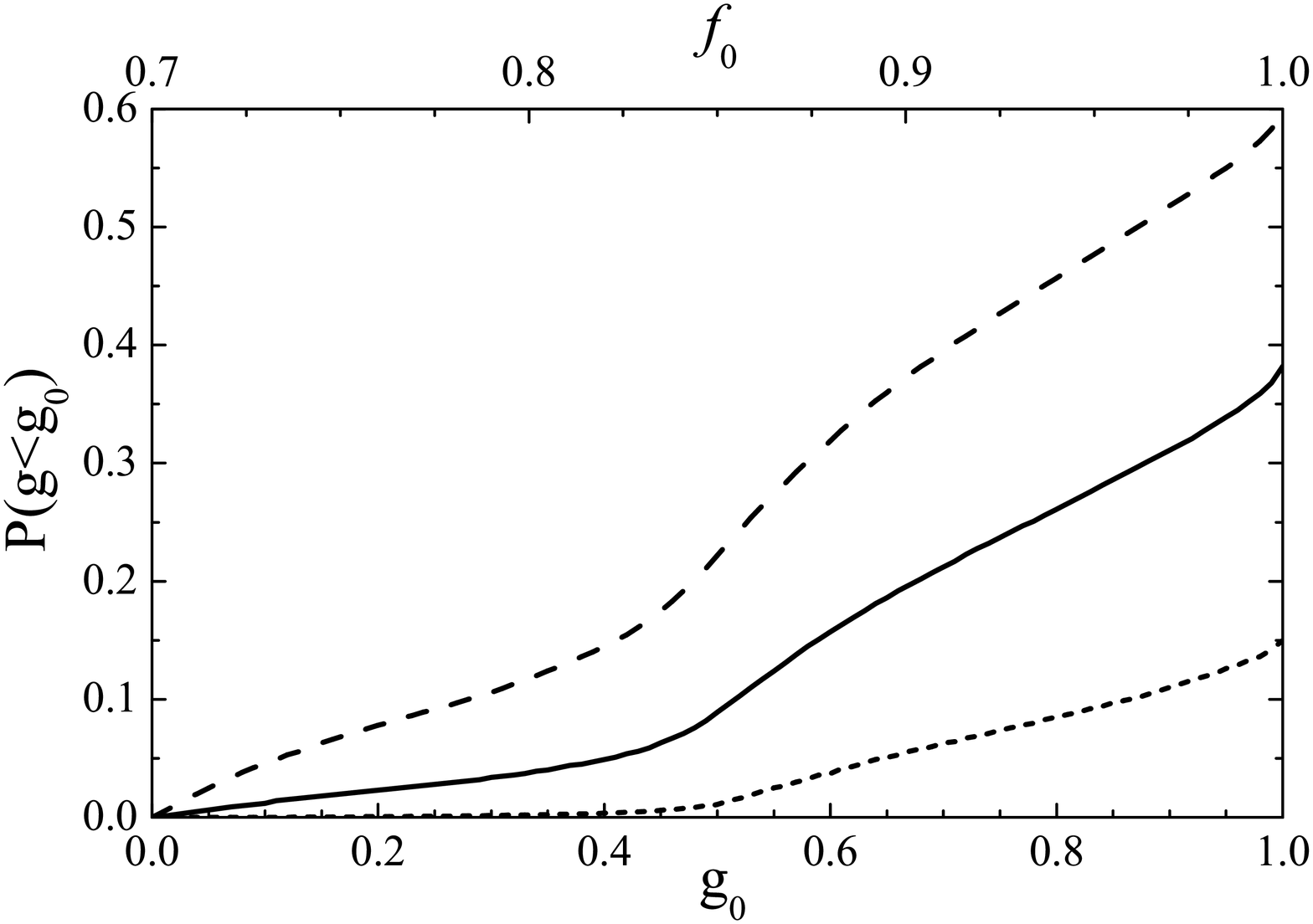}
\caption{
{\it Left panel}: Three types of cloud sizes,
$l_c$, distribution.
Short-dashed  curve is the uniform distribution
within an interval of
$0.2  \leq l_c \leq 15~{\rm pc}$,
solid curve is the lognormal distribution
on a condition that the values $l_c$
with 90$\%$ probability
belong to the same interval
$0.2  \leq l_c \leq 15~{\rm pc}$,
long-dashed curve is the lognormal distribution
on a condition that the most probable
value is $l_c=0.2$~pc (see text).
{\it Right panel}: the cumulative 
distribution function $P(g<g_0)$ 
--- the probability
to detect absorption systems 
in a QSO spectrum 
(represented at $z_q=2.57$ and $z_c=2.33$)
with  the geometrical coverage factor $g < g_0$. 
The upper argument, $f_0$
(flux coverage factor), 
determined by a  
scaling $f_0=0.3 g_0 + 0.7$
in such a way that the
cumulative 
distribution 
functions $P(f<f_0)$ 
remains unalterable.
Types of lines are the same as in the {\it left panel}.
}
\label{Distr}
\end{figure*}

Actually,
Fig.~\ref{LumBLR} shows the distributions
of QSO (DLA) luminosities
at $\lambda = 1450$\,\AA\ (left panel) and
$p_{qso}(r)$ calculated
for BLR radii (right panel).
One can see that the lognormal approximation
\begin{equation}
p_{qso}(r) = {1 \over \sqrt{2 \pi} w}
             {e^{(ln {r \over r_0})^2 \over 2 w^2} \over r},
\label{p_r}
\end{equation}
where $w=0.426 \pm 0.003$ and $r_0 = (0.1442 \pm 0.0004)$~pc
obtained by the least square method,
proves to be good.

The  situation with a distribution $p_{cloud}(l_c)$ for
H$_2$ molecular clouds
is still more uncertain. 
Absorption systems of H$_2$ detected at high redshifts 
are referred to the cold neutral medium, which has a lower limit
of the number density about
$1 \div 10$~cm$^{-3}$ (e.g. see \citealt{Liszt2002}). 
It gives 
$l_c < 3 \div 30$~pc
for typical measurable column densities
$\sim 10^{19} \div 10^{20}$~cm$^{-2}$. 
On the other hand,
the analysis of H$_2$ absorption 
system in the spectrum of Q1232+082 (\citealt{Balashev2011})
yields $l_c \sim 0.2$~pc.
Therefore we employ for our estimation
two modeled distributions:
\noindent
(1) the uniform distribution within an interval $[0.2,\, 15.0]$~pc
\begin{equation}
p_{cloud}^{(1)}(l_c) = {1 \over b-a} \Theta(l_c-a) \Theta(b-l_c),
\label{p1}
\end{equation}
where $a=0.2$~pc and $b=15.0$~pc. 
One can expect that this model gives
overestimated sizes of the clouds and consequently
underestimated values of $P(g > g_0)$.
\noindent
(2) The lognormal distribution
\begin{equation}
p_{cloud}^{(2)}(l_c) = {1 \over \sqrt{2 \pi} u}
             {e^{(ln {l_c \over l_0})^2 \over 2 u^2} \over l_c},
\label{p2}
\end{equation}
where parameters $u$ and $l_0$ are choosing also for two
cases. The first one is $u=1.2$ and $l_0=0.8$~pc, it corresponds
to the fixed maximum of the distribution
Eq.~(\ref{p2}) or to the most probable size at $l_c=0.2$.
The second one satisfies to the condition that
90$\%$ clouds fall into the interval $[0.2, 15.0]$~pc.
In that case $u=1.3$ and $l_0=2.0$~pc.

The left panel in Fig.~\ref{Distr} demonstrates three
types of the PDF for the cloud sizes
$l_c$ defined by  Eqs.~(\ref{p1}) and  
(\ref{p2}) with two sets of parameters $u$ and $l_0$.
The right panel shows the integral probability
$P(g<g_0)$ as functions of $g_0$
calculated with the use of
Eqs.~(\ref{P(g)}),
(\ref{p_delta}), and  (\ref{p_kappa})
at all three  types of distribution 
$p_{cloud}(l_c=2\kappa r)$ 
discussed above. Let us emphasize
the strong dependence of the probability distribution
on the distribution of cloud sizes. Nevertheless,
even the most conservative estimation of the
probability to reveal $g< 0.98$ proves to be $\approx 15\%$,
where $g_0=0.98$ corresponds to the minimal measurable
noncoverage factor: $1-g_0 \geq 0.02$  
(\citealt{Klimenko2015}).

\subsection{Flux coverage factor}
\label{Distr-Flux}
In contrast with the oversimplified geometrical
coverage factor $g$ 
one can introduce
the factor $f$ as a ratio of
radiation fluxes,
which is more  relevant
to a spectral analyses:
\begin{equation}
f={I_{cov} \over I_{qso}},
\label{f}
\end{equation}
where $I_{cov}$ is the flux of the quasar radiation
passed through the cloud, $I_{qso}$ is
the total flux.
This value may be called as a {\it flux
coverage factor}. In the case of uniform
distribution of the flux over a solid angle
$\Omega_q$ we have $I_{cov} \propto \Omega_{cov}$,
$I_{qso} \propto \Omega_{q}$ and the flux factor $f$
would be equal to $g$.
\begin{figure*}
\centering
\includegraphics[width=\textwidth]{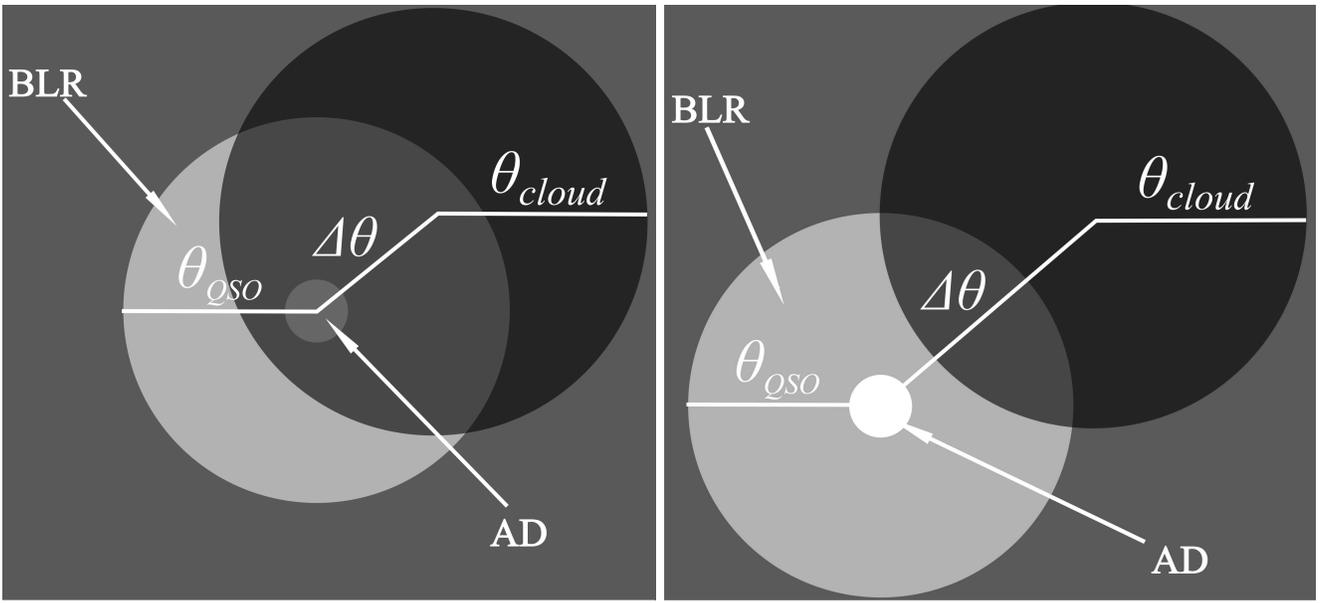}
\caption{
Model representation of
radiation-flux nonuniformity.
{\it Left panel}:
the accretion disk of QSO is covered by a
cloud totally, $\Delta \theta < \theta_{cloud}$\, 
($\delta < \rho$); 
{\it right panel}: the accretion disk of QSO
is uncovered by a cloud, 
$\Delta \theta > \theta_{cloud}$\,
($\delta > \rho$).
The flux passing by the cloud contains the whole flux
of the accretion disk and more than a half
of flux emitted by the BLR.
In both case the BLR is covered only partially,
but the 
case represented on the right panel
is unlikely to be detected
(see  Section\,\ref{Distr-Geom} for details).
}
\label{Fig_flux}
\end{figure*}

Let us
regard, some arbitrary, that
the flux 
from  the AD
characterized by a continuum spectrum
constitutes about 70$\%$
of the whole intensity.
The rest 30$\%$
of the flux originates
from the BLR (Fig.~\ref{Fig_flux})
and represents a set of broad 
emission lines which
also raises a level of continuum
in some parts of the spectrum\footnote{
The fact that a continuum source can be fully associated 
with the AD is proved by observations of rapid continuum 
variability (see \citealt{Peterson1993} 
and references therein)}.

If the AD is covered totally  
but the BLR is covered partially
({\it left panel} in  Fig.~\ref{Fig_flux}),
the partial coverage effect ($f<1$) will be detected 
only 
in those
absorption lines which are located 
around the top
of QSO emission lines.  
Outside the wide emission lines 
the partial coverage 
is not detectable, that
is equivalent to $f = 1$.
Within the regions of emission lines  
the full covering of 
the AD flux yields only a part
of the covering   
factor $f=0.7$ and the 
rest part of $f$ is provided by
the geometrical coverage factor $g$ 
related solely to the BLR. 
As a result we assume that
the flux coverage factor $f$ is determined 
by a simple equation:
\begin{equation}
f = 0.7 + 0.3 g .
\label{f_g}
\end{equation}

Then, the calculation of the CDF $P(f<f_0)$ is 
reduced
to calculation of $P(g<(f_0-0.7)/0.3)$, i.e. to pure
geometrical approach. 
A scaling $f_0=0.7 + 0.3 g_0$ applied to the upper 
horizontal axis in the right panel of  Fig.~\ref{Distr} allows 
one to transform the dependencies  on $g$ --- $P(g<g_0)$ 
into the dependencies on $f$ ---
$P(f<f_0)$.

The most conservative estimation
of the probability to find a partial coverage
at the minimal level  of the
noncoverage factor  $1-f_0 \geq 0.02$ or
$f_0 \leq 0.98$  (e.g. \citealt{Klimenko2015})
yields 11$\%$ (lower curve in the 
right panel of Fig.~\ref{Distr}).
Note, that to detect the residual flux at the bottom 
of lines (which typically include several pixels) 
at the level $\sim 2\,\%$ it is necessary to 
obtain high resolution spectra with 
high signal to noise ratio, ${\rm S/N\sim30-50}$. 
Spectra registered
with less S/N 
allow to detect only
higher minimal level, e.g.
$1-f_0 \ga 0.1$ or $f_0 \la 0.9$.
It gives the lowest level 
of the probability: $\sim 5\%$.
However, we suppose that the middle solid curve in
Fig.~\ref{Distr} is more realistic. It gives
the probability estimation of the order of $15\%$
to reveal the partial coverage with $f < 0.9$.
Thus, the effect of partial coverage 
of QSOs by intervening H$_2$ ALSs characterized by
noticeable probability to be revealed
at the spectra analyses.

\subsection{Dependencies of $f$-distribution on redshifts}
\label{redshift}
\begin{figure}
\centering
\includegraphics[width=\columnwidth]{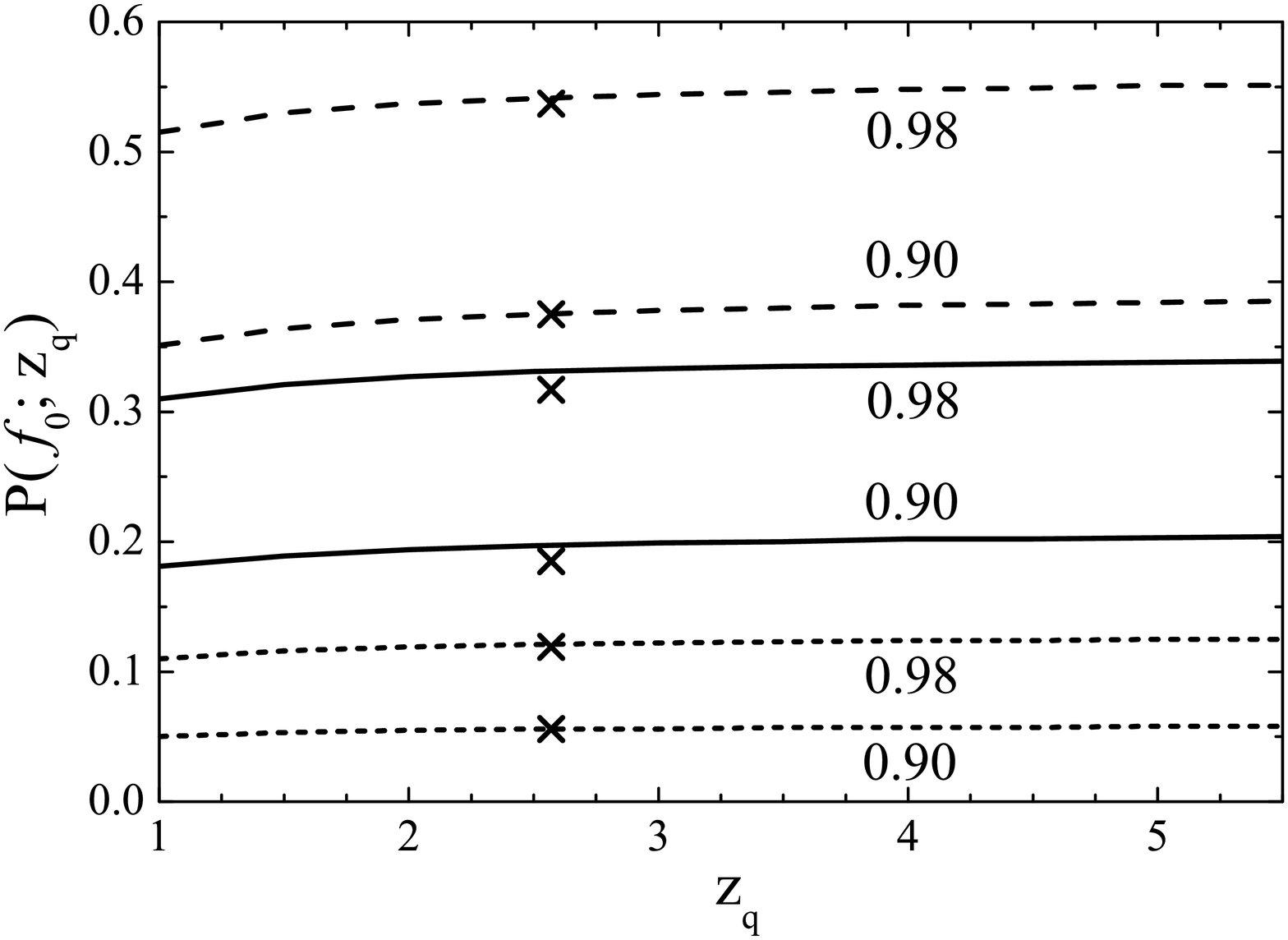}
\caption{
The cumulative distribution
$P(f_0;\, z_q)=P(f<f_0; z_q)$ (see text)
as a function of QSO redshift, $z_q$, when
a cloud redshift $z_c=z_c(z_q)$ is
maximally remote from $z_q$
(see  Eq.~\protect{(\ref{zc})}).
Types of lines are the  same as in left  
panel of Fig.~\ref{Distr}. 
Each pair of the curves corresponds to the same
distribution of $l_c$; the higher curve
of a pair is plotted at 
$f_0=0.98$, while the lower one -- at $f_0=0.90$
(numbers near the curves). 
Crossed points correspond to the special case of $P(f_0;\ z_q)$
calculated for Q1232+082 at $z_q=2.57$ and the H$_2$-cloud
at $z_c=2.33$ registered in its spectrum.
}
\label{P(zq)}
\end{figure}

Let us consider dependencies 
of the probability distribution 
on the redshifts $z_q$ and $z_c$. 
Taking into account dependence
of the distribution $p_{\delta}(\delta| \kappa)$
on  $z_q$ and $z_c$, 
defined by Eqs.~(\ref{p_delta}) and (\ref{rho_kappa}),
one can modify 
Eq.~\ref{P(g)} and obtain CDF as
\begin{multline}
P(f<f_0; z_q, z_c) = \\
\int_0^\infty \int_0^{\kappa[(f_0-0.7)/0.3,\,\delta, z_q, z_c]}  
p_{\delta}(\delta| \kappa, z_q, z_c) p_{\kappa}(\kappa)\,
{\rm d} \delta\,  {\rm d}\kappa,
\label{P(f)}  
\end{multline}
where according to Eq.~(\ref{f_g}) 
${f_0}$ varies between $0.7$ and $1.0$, and the 
function $\kappa$ is given in Eqs. (\ref{k_a}) and (\ref{k_b}).

Firstly, the dependence of the CDF 
on redshifts are expected to be small, 
since the angular diameter 
distance (see Eq.~(\ref{DA})) 
weakly depends on the redshifts
within an interval $z_c,\ z_q\ \sim 2-4$.
Let us consider an H$_2$-cloud, 
which is remote from the QSO 
by the maximal distance 
consistent with the very
possibility to detect a set of 
H$_2$ absorption lines in surrounding
of a broad emission line
and consequently 
to detect the partial coverage.

One can roughly formulate this 
condition of maximal removing as 
$(1+z_c)\,1094\text{\,\AA} > (1+z_q)\,977\text{\,\AA}$, 
where 1094\,\AA\, 
is wavelength of L1-0 P(1) transition and
977.02\,\AA\,  corresponds to the 
C\,{\sc iii} prominent emission line
in the restframe (see Fig.~\ref{spec}),
since most of saturated 
H$_2$ absorption lines 
are blueward in relation to 1094\,\AA\,  
and all prominent emission lines 
of QSO spectra 
are redward in relation to 977.02\,\AA.
Note that we conservatively take L1-0 P(1) 
line as a boundary, since the most long-wavelength 
H$_2$ transitions correspond to L0-0 band at 
$\lambda \sim 1108-1110$\,\AA\ (see Fig.~\ref{spec}) 
frequently appears as
a  unsaturated  absorption lines
and can not be used for confident  detection of the
partial coverage.
Therefore,
to obtain partial coverage of the QSO BLR by  
a H$_2$-cloud one needs that
redshifts of H$_2$ ALSs
obey the condition 
\begin{equation}
0.89 \,z_q - 0.11 \leq z_c\leq z_q,
\label{zc}
\end{equation}  
where we put $977/1094 \simeq 0.89$.
The right part of condition just indicates
that H$_2$ ALS is 
located between QSO and observer.  

Fig.~\ref{P(zq)} demonstrates 
dependencies $P(f_0;\, z_q)=P(f<f_0;\, z_q, z_c(z_q))$
on $z_q$ at two fixed $f_0=0.98$ and $0.90$ 
and $z_c(z_q)$ determined 
by  minimal values of Eq.~(\ref{zc}).
One can see that the probability 
to reveal the flux coverage factor
$0.7 < f < f_0$ weakly depends on
$z_q$. Moreover, the probability $P(f_0, z_q, z_c)$ 
weakly depends also on $z_c$. 
Actually, it is demonstrated by crossed
points in Fig.~\ref{P(zq)} calculated for
the QSO\,1232+082 ($z_q=2.57$) 
whose spectrum contains
the H$_2$ absorption system at $z_c=2.33$.
The crossed points lie closely to the corresponding
curves $P(f_0; z_q)$, 
while the lower boundary 
$z_c$ at $z_q=2.57$ according to Eq.~(\ref{zc}) is 
$z_c \geq 2.17$, i.e. the difference does not affect
noticeably the value $P(f_0; z_q)$. 
Thus we can disregard 
approximately the dependence
$P(f_0;\,  z_q)$ on $z_c$.

\section{Notes on observations of the partial 
coverage effect in QSO spectra}
\label{part_cov}

\begin{figure*}
	\centering
   \includegraphics[width = \textwidth]{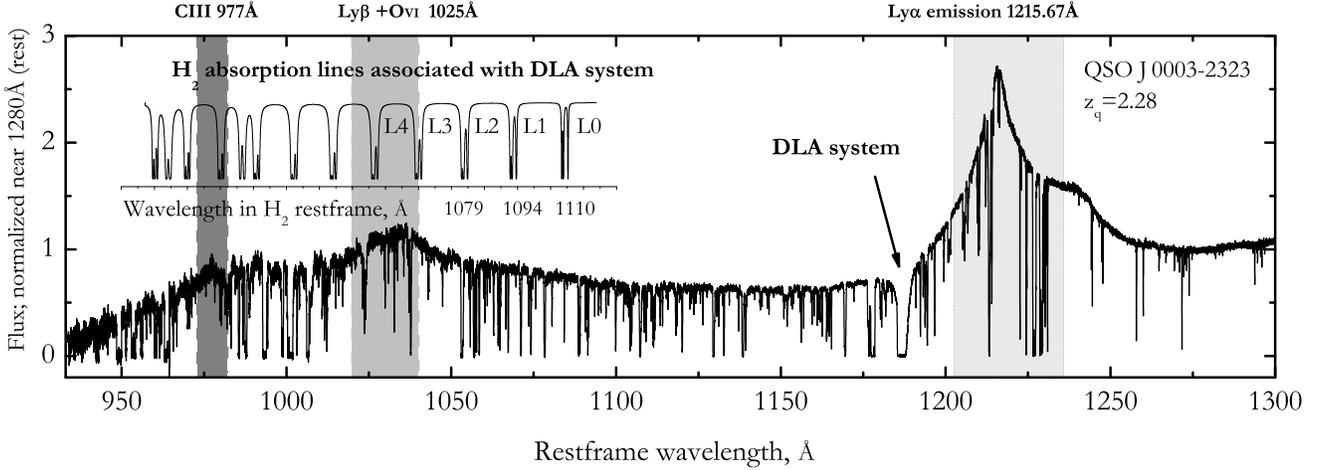}
	\caption{ 
	Spectrum of quasar J\,0003-2323 at $z_q=2.28$
	shown in its own restframe. There are marked Ly$_{\alpha}$ emission line, 
	DLA system at $\lambda_{\text{DLA}} = 1215\text{\AA} (1+z_c)/(1+z_q)$,
	broad emission lines O\,{\sc iv}, Ly$_{\beta}$ and C\,{\sc iii}, 
	Grey vertical bands indicate the wavelength
	regions in surroundings of the broad 
	emission lines where the
	effect of partial coverage could be mostly pronounced.
	In the upper left part of the figure it is shown also
	the series of typical absorption bands
	of the H$_2$ molecules. The spectrum is not properly flux 
	calibrated, since it was obtained using Ultra Violet Echelle 
	Spectrograph at VLT. For data see the footnote 
        in the Section~\ref{part_cov}.
}
\label{spec}
\end{figure*}

\begin{figure}
	\includegraphics[width = \columnwidth]{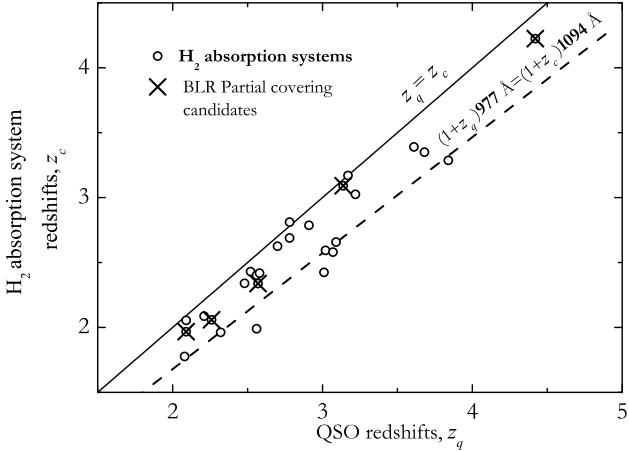}
	\caption{
   Known high-redshifted H$_2$ absorption systems 
	on the $z_q-z_c$ plane.
	Empty circles indicate absorption systems, 
	where partial coverage 
	has not been detected up to now (22 objects); 
	crosses mark systems for which a preliminary analysis 
	evidences in favour of an occurrence of 
	partially covered BLRs (5 objects) in QSOs spectra. 
	The upper solid line corresponds to $z_c=z_q$ and
	determines the highest positions of H$_2$ systems 
	on the $z_q-z_c$ plane; 
	the dashed line is $z_c = 0.89z_q - 0.11$ 
	(see Eq.~(\ref{zc}))
	and gives the lowest $z_c$ at fixed $z_q$ 
	when partial coverage by intervening 
	cloud can be detected. 
        The data was taken from \citet{Ivanchik2015}.
}
\label{zczq}
\end{figure}

As an illustration
Fig.~\ref{spec} displays  
typical spectrum of the QSO in the restframe.
Specifically we used the quasar spectrum 
J\,0003-2323 at $z_q=2.28$. 
It was observed under program 166.A-0106 (PI: Jacqueline Bergeron). 
The data are in free access 
from the ESO database\footnote{http://archive.eso.org}.
One can see strong Ly$_{\alpha}$ emission line, 
a number of absorption lines of the Lyman-alpha forest 
imprinted in a continuum, damped  Lyman-alpha absorption  
(DLA) system at $\lambda_{\text{DLA}} = 1215\text{\,\AA} (1+z_c)/(1+z_q)$, 
broad emission lines O\,{\sc iv}, Ly$_{\beta}$ and C\,{\sc iii}, 
and schematically a set of 
H$_2$-molecular absorption bands. 
We regard (see Section~\ref{Distr-Geom}) 
that  H$_2$-bearing clouds are supposed 
to be associated with DLA systems. 

Fig.~\ref{zczq} shows the location 
of known
H$_2$ absorption 
systems at high redshifts on the $z_q - z_c$ plane. 
All H$_2$ systems capable to cause the
partial coverage of QSO BLRs can be placed mainly 
between two linear bounds determined by Eq.~(\ref{zc}). 
Fig.~\ref{zczq} represents also 
preliminary results 
of a systematic search of H$_2$-systems with traces of 
the partial coverage effect (Klimenko et al, in prep.).
All five candidates 
marked by crosses in Fig.~\ref{zczq}
have to be
verified in further analysis.

\section{Conclusions}
\label{concl}
In this paper we estimate the probability of an incomplete
coverage of the QSO BLR by the absorbing H$_2$ molecular
cloud. The key value of our analysis is a coverage factor
$f$ -- a ratio of the flux of radiation 
which goes through a cloud
to the total flux could have registered from the QSO
without absorption by the cloud. 
We calculate the probability to obtain different
coverage factors at given cosmological 
redshifts $z_q$ and $z_c$
of the QSO and cloud, respectively, 
employing the distribution of BLR sizes
based on the data of the SDSS\, DR9, and
a few model distributions of cloud sizes.

According to Fig.~\ref{Distr} the most conservative
assessment of the probability to reveal the factor
$f < 0.98$ (observational minimum of noncoverage 
factor $1-f \geq 0.02$) is $11\%$. Thus
one can expect at least three cases with $f < 0.98$
from  27~H$_2$ absorption systems detected in QSO spectra
up to now. The preliminary results of  
the systematic search of the partial coverage effect
in quasar spectra with imprinted intervening 
H$_2$ absorption systems  give five cases, 
that is consistent with the prediction. 

Systematic search for the partial coverage effect 
in the H$_2$ absorption systems will allow to obtain 
restrictions on characteristic scales of the H$_2$ clouds.  
For instance, relatively high
frequency of revealing the effects of incomplete coverage
with $0.7 < f < 1.0$ would mean that cloud sizes are comparable
with the BLR sizes, i.e. $\sim 0.2 - 0.3$~pc. Relatively low
frequency of such effects in QSO spectra  would mean
that the typical cloud  size  essentially exceeds the
BLR size as it was widely accepted previously.

In this paper we consider a simple model, 
where absorption and emission regions 
are represented by spherical shape clouds. 
In case of asymmetrical and/or patchy structure 
of the absorption as well as emission 
regions the situation can be sufficiently 
complicated by increasing the number of parameters 
and uncertainties.  However, it seems that estimate 
of the probability of revealing partial 
coverage -- the main result of this paper, 
will not significantly changed. 
Our understanding of the structure, 
distribution of sizes and physical conditions 
of cold interstellar medium at high redshift 
would be constrained by
the further observations 
and extension
of the statistics of the BLRs partial 
coverage by H$_2$ absorption clouds.

\acknowledgments
This work is supported 
by the Russian Science Foundation, 
grant 14-12-00955.

\bibliographystyle{spr-mp-nameyear-cnd}     
\bibliography{DDO_clouds_bib}                

\begin{thebibliography}{26}
\ifx \bisbn   \undefined \def \bisbn  #1{ISBN #1}\fi
\ifx \binits  \undefined \def \binits#1{#1} \fi
\ifx \bauthor  \undefined \def \bauthor#1{#1} \fi
\ifx \batitle  \undefined \def \batitle#1{#1} \fi
\ifx \bjtitle  \undefined \def \bjtitle#1{#1}\fi
\ifx \bvolume  \undefined \def \bvolume#1{\textbf{#1}}\fi
\ifx \byear  \undefined \def \byear#1{#1} \fi
\ifx \bissue  \undefined \def \bissue#1{#1} \fi
\ifx \bfpage  \undefined \def \bfpage#1{#1} \fi
\ifx \blpage  \undefined \def \blpage #1{#1} \fi
\ifx \burl  \undefined \def \burl#1{\textsf{#1}} \fi
\ifx \doiurl  \undefined \def \doiurl#1{\textsf{#1}} \fi
\ifx \betal  \undefined \def \betal{\textit{et al.}} \fi
\ifx \binstitute  \undefined \def \binstitute#1{#1} \fi
\ifx \binstitutionaled  \undefined \def \binstitutionaled#1{#1} \fi
\ifx \bctitle  \undefined \def \bctitle#1{#1} \fi
\ifx \beditor  \undefined \def \beditor#1{#1} \fi
\ifx \bpublisher  \undefined \def \bpublisher#1{#1} \fi
\ifx \bbtitle  \undefined \def \bbtitle#1{#1} \fi
\ifx \bedition  \undefined \def \bedition#1{#1} \fi
\ifx \bseriesno  \undefined \def \bseriesno#1{#1} \fi
\ifx \blocation  \undefined \def \blocation#1{#1} \fi
\ifx \bsertitle  \undefined \def \bsertitle#1{#1} \fi
\ifx \bsnm \undefined \def \bsnm#1{#1} \fi
\ifx \bsuffix \undefined \def \bsuffix#1{#1} \fi
\ifx \bparticle \undefined \def \bparticle#1{#1} \fi
\ifx \barticle \undefined \def \barticle#1{#1} \fi
\ifx \bconfdate \undefined \def \bconfdate #1{#1} \fi
\ifx \botherref \undefined \def \botherref #1{#1} \fi
\ifx \url \undefined \def \url#1{\textsf{#1}} \fi
\ifx \bchapter \undefined \def \bchapter#1{#1} \fi
\ifx \bbook \undefined \def \bbook#1{#1} \fi
\ifx \bcomment \undefined \def \bcomment#1{#1} \fi
\ifx \oauthor \undefined \def \oauthor#1{#1} \fi
\ifx \citeauthoryear \undefined \def \citeauthoryear#1{#1} \fi
\ifx \endbibitem  \undefined \def \endbibitem {}\fi
\ifx \bconflocation  \undefined \def \bconflocation#1{#1} \fi
\ifx \arxivurl  \undefined \def \arxivurl#1{\textsf{#1}} \fi

\bibitem[\protect\citeauthoryear{{Abdo} et~al.}{2010}]{Abdo2010}
\begin{barticle}
\bauthor{\bsnm{{Abdo}}, \binits{A.A.}},
\bauthor{\bsnm{{Ackermann}}, \binits{M.}},
\bauthor{\bsnm{{Ajello}}, \binits{M.}},
\bauthor{\bsnm{{Axelsson}}, \binits{M.}},
\bauthor{\bsnm{{Baldini}}, \binits{L.}},
\bauthor{\bsnm{{Ballet}}, \binits{J.}},
\bauthor{\bsnm{{Barbiellini}}, \binits{G.}},
\bauthor{\bsnm{{Bastieri}}, \binits{D.}},
\bauthor{\bsnm{{Baughman}}, \binits{B.M.}},
\bauthor{\bsnm{{Bechtol}}, \binits{K.}},
\bauthor{\bparticle{et} \bsnm{al.}}:
\bjtitle{\nat}
\bvolume{463},
\bfpage{919}
(\byear{2010})
\end{barticle}
\endbibitem

\bibitem[\protect\citeauthoryear{{Ahn} and et~al.}{2012}]{Ahn2012}
\begin{barticle}
\bauthor{\bsnm{{Ahn}}, \binits{C.P.}},
\bauthor{\bparticle{et} \bsnm{al.}}:
\bjtitle{\apjs}
\bvolume{203},
\bfpage{21}
(\byear{2012})
\end{barticle}
\endbibitem

\bibitem[\protect\citeauthoryear{{Balashev} et~al.}{2011}]{Balashev2011}
\begin{barticle}
\bauthor{\bsnm{{Balashev}}, \binits{S.A.}},
\bauthor{\bsnm{{Petitjean}}, \binits{P.}},
\bauthor{\bsnm{{Ivanchik}}, \binits{A.V.}},
\bauthor{\bsnm{{Ledoux}}, \binits{C.}},
\bauthor{\bsnm{{Srianand}}, \binits{R.}},
\bauthor{\bsnm{{Noterdaeme}}, \binits{P.}},
\bauthor{\bsnm{{Varshalovich}}, \binits{D.A.}}:
\bjtitle{\mnras}
\bvolume{418},
\bfpage{357}
(\byear{2011})
\end{barticle}
\endbibitem

\bibitem[\protect\citeauthoryear{{Blackburne} et~al.}{2011}]{Blackburne2011}
\begin{barticle}
\bauthor{\bsnm{{Blackburne}}, \binits{J.A.}},
\bauthor{\bsnm{{Pooley}}, \binits{D.}},
\bauthor{\bsnm{{Rappaport}}, \binits{S.}},
\bauthor{\bsnm{{Schechter}}, \binits{P.L.}}:
\bjtitle{\apj}
\bvolume{729},
\bfpage{34}
(\byear{2011})
\end{barticle}
\endbibitem

\bibitem[\protect\citeauthoryear{{Chelouche} and
  {Daniel}}{2012}]{Chelouche2012}
\begin{barticle}
\bauthor{\bsnm{{Chelouche}}, \binits{D.}},
\bauthor{\bsnm{{Daniel}}, \binits{E.}}:
\bjtitle{\apj}
\bvolume{747},
\bfpage{62}
(\byear{2012})
\end{barticle}
\endbibitem

\bibitem[\protect\citeauthoryear{{Ibata} et~al.}{1999}]{Ibata1999}
\begin{barticle}
\bauthor{\bsnm{{Ibata}}, \binits{R.A.}},
\bauthor{\bsnm{{Lewis}}, \binits{G.F.}},
\bauthor{\bsnm{{Irwin}}, \binits{M.J.}},
\bauthor{\bsnm{{ Leh{\'a}r}}, \binits{J.}},
\bauthor{\bsnm{{Totten}}, \binits{E.J.}}:
\bjtitle{\aj}
\bvolume{118},
\bfpage{1922}
(\byear{1999})
\end{barticle}
\endbibitem

\bibitem[\protect\citeauthoryear{{Ivanchik} et~al.}{2010}]{Ivanchik2010}
\begin{barticle}
\bauthor{\bsnm{{Ivanchik}}, \binits{A.V.}},
\bauthor{\bsnm{{Petitjean}}, \binits{P.}},
\bauthor{\bsnm{{Balashev}}, \binits{S.A.}},
\bauthor{\bsnm{{Srianand}}, \binits{R.}},
\bauthor{\bsnm{{Varshalovich}}, \binits{D.A.}},
\bauthor{\bsnm{{Ledoux}}, \binits{C.}},
\bauthor{\bsnm{{Noterdaeme}}, \binits{P.}}:
\bjtitle{\mnras}
\bvolume{404},
\bfpage{1583}
(\byear{2010})
\end{barticle}
\endbibitem

\bibitem[\protect\citeauthoryear{{Ivanchik} et~al.}{2015}]{Ivanchik2015}
\begin{barticle}
\bauthor{\bsnm{{Ivanchik}}, \binits{A.V.}},
\bauthor{\bsnm{{Balashev}}, \binits{S.A.}},
\bauthor{\bsnm{{Varshalovich}}, \binits{D.A.}},
\bauthor{\bsnm{{Klimenko}}, \binits{V.V.}}:
\bjtitle{Astronomy Reports}
\bvolume{59},
\bfpage{100}
(\byear{2015})
\end{barticle}
\endbibitem

\bibitem[\protect\citeauthoryear{{Jim{\'e}nez-Vicente}
  et~al.}{2012}]{JimenezVicente2012}
\begin{barticle}
\bauthor{\bsnm{{Jim{\'e}nez-Vicente}}, \binits{J.}},
\bauthor{\bsnm{{Mediavilla}}, \binits{E.}},
\bauthor{\bsnm{{Mu{\~n}oz}}, \binits{J.A.}},
\bauthor{\bsnm{{Kochanek}}, \binits{C.S.}}:
\bjtitle{\apj}
\bvolume{751},
\bfpage{106}
(\byear{2012})
\end{barticle}
\endbibitem

\bibitem[\protect\citeauthoryear{{Kanekar} et~al.}{2009}]{Kanekar2009}
\begin{barticle}
\bauthor{\bsnm{{Kanekar}}, \binits{N.}},
\bauthor{\bsnm{{Lane}}, \binits{W.M.}},
\bauthor{\bsnm{{Momjian}}, \binits{E.}},
\bauthor{\bsnm{{Briggs}}, \binits{F.H.}},
\bauthor{\bsnm{{Chengalur}}, \binits{J.N.}}:
\bjtitle{\mnras}
\bvolume{394},
\bfpage{61}
(\byear{2009})
\end{barticle}
\endbibitem

\bibitem[\protect\citeauthoryear{{Kaspi} et~al.}{2005}]{Kaspi2005}
\begin{barticle}
\bauthor{\bsnm{{Kaspi}}, \binits{S.}},
\bauthor{\bsnm{{Maoz}}, \binits{D.}},
\bauthor{\bsnm{{Netzer}}, \binits{H.}},
\bauthor{\bsnm{{Peterson}}, \binits{B.M.}},
\bauthor{\bsnm{{Vestergaard}}, \binits{M.}},
\bauthor{\bsnm{{Jannuzi}}, \binits{B.T.}}:
\bjtitle{\apj}
\bvolume{629},
\bfpage{61}
(\byear{2005})
\end{barticle}
\endbibitem

\bibitem[\protect\citeauthoryear{{Kaspi} et~al.}{2007}]{Kaspi2007}
\begin{barticle}
\bauthor{\bsnm{{Kaspi}}, \binits{S.}},
\bauthor{\bsnm{{Brandt}}, \binits{W.N.}},
\bauthor{\bsnm{{Maoz}}, \binits{D.}},
\bauthor{\bsnm{{Netzer}}, \binits{H.}},
\bauthor{\bsnm{{Schneider}}, \binits{D.P.}},
\bauthor{\bsnm{{Shemmer}}, \binits{O.}}:
\bjtitle{\apj}
\bvolume{659},
\bfpage{997}
(\byear{2007})
\end{barticle}
\endbibitem

\bibitem[\protect\citeauthoryear{{Kayser} et~al.}{1997}]{Kayser1997}
\begin{barticle}
\bauthor{\bsnm{{Kayser}}, \binits{R.}},
\bauthor{\bsnm{{Helbig}}, \binits{P.}},
\bauthor{\bsnm{{Schramm}}, \binits{T.}}:
\bjtitle{\aap}
\bvolume{318},
\bfpage{680}
(\byear{1997})
\end{barticle}
\endbibitem

\bibitem[\protect\citeauthoryear{{Klimenko} et~al.}{2015}]{Klimenko2015}
\begin{barticle}
\bauthor{\bsnm{{Klimenko}}, \binits{V.V.}},
\bauthor{\bsnm{{Balashev}}, \binits{S.A.}},
\bauthor{\bsnm{{Ivanchik}}, \binits{A.V.}},
\bauthor{\bsnm{{Ledoux}}, \binits{C.}},
\bauthor{\bsnm{{Noterdaeme}}, \binits{P.}},
\bauthor{\bsnm{{Petitjean}}, \binits{P.}},
\bauthor{\bsnm{{Srianand}}, \binits{R.}},
\bauthor{\bsnm{{Varshalovich}}, \binits{D.A.}}:
\bjtitle{\mnras}
\bvolume{448},
\bfpage{280}
(\byear{2015})
\end{barticle}
\endbibitem

\bibitem[\protect\citeauthoryear{{Kulkarni} and {Loeb}}{2012}]{Kulkarni2012}
\begin{barticle}
\bauthor{\bsnm{{Kulkarni}}, \binits{G.}},
\bauthor{\bsnm{{Loeb}}, \binits{A.}}:
\bjtitle{\mnras}
\bvolume{422},
\bfpage{1306}
(\byear{2012})
\end{barticle}
\endbibitem

\bibitem[\protect\citeauthoryear{{Ledoux} et~al.}{1998}]{Ledoux1998}
\begin{barticle}
\bauthor{\bsnm{{Ledoux}}, \binits{C.}},
\bauthor{\bsnm{{Theodore}}, \binits{B.}},
\bauthor{\bsnm{{Petitjean}}, \binits{P.}},
\bauthor{\bsnm{{Bremer}}, \binits{M.N.}},
\bauthor{\bsnm{{Lewis}}, \binits{G.F.}},
\bauthor{\bsnm{{Ibata}}, \binits{R.A.}},
\bauthor{\bsnm{{Irwin}}, \binits{M.J.}},
\bauthor{\bsnm{{Totten}}, \binits{E.J.}}:
\bjtitle{\aap}
\bvolume{339},
\bfpage{77}
(\byear{1998})
\end{barticle}
\endbibitem

\bibitem[\protect\citeauthoryear{{Liszt}}{2002}]{Liszt2002}
\begin{barticle}
\bauthor{\bsnm{{Liszt}}, \binits{H.}}:
\bjtitle{\aap}
\bvolume{389},
\bfpage{393}
(\byear{2002})
\end{barticle}
\endbibitem

\bibitem[\protect\citeauthoryear{{Noterdaeme} and
  et~al.}{2012}]{Noterdaeme2012}
\begin{barticle}
\bauthor{\bsnm{{Noterdaeme}}, \binits{P.}},
\bauthor{\bparticle{et} \bsnm{al.}}:
\bjtitle{\aap}
\bvolume{547},
\bfpage{1}
(\byear{2012})
\end{barticle}
\endbibitem

\bibitem[\protect\citeauthoryear{{Ofengeim} et~al.}{2013}]{Ofengeim2013}
\begin{barticle}
\bauthor{\bsnm{{Ofengeim}}, \binits{D.D.}},
\bauthor{\bsnm{{Ivanchik}}, \binits{A.V.}},
\bauthor{\bsnm{{Kaminker}}, \binits{A.D.}},
\bauthor{\bsnm{{Balashev}}, \binits{S.A.}}:
\bjtitle{J. Phys.: Conf. Ser.}
\bvolume{461}(\bissue{1}),
\bfpage{012046}
(\byear{2013})
\end{barticle}
\endbibitem

\bibitem[\protect\citeauthoryear{{Peterson}}{1993}]{Peterson1993}
\begin{barticle}
\bauthor{\bsnm{{Peterson}}, \binits{B.M.}}:
\bjtitle{\pasp}
\bvolume{105},
\bfpage{247}
(\byear{1993})
\end{barticle}
\endbibitem

\bibitem[\protect\citeauthoryear{{Petitjean} et~al.}{1994}]{Petitjean1994}
\begin{barticle}
\bauthor{\bsnm{{Petitjean}}, \binits{P.}},
\bauthor{\bsnm{{Rauch}}, \binits{M.}},
\bauthor{\bsnm{{Carswell}}, \binits{R.F.}}:
\bjtitle{\aap}
\bvolume{291},
\bfpage{29}
(\byear{1994})
\end{barticle}
\endbibitem

\bibitem[\protect\citeauthoryear{{Planck Collaboration}
  et~al.}{2014}]{Planck2013XVI}
\begin{barticle}
\bauthor{\bsnm{{Planck Collaboration}}},
\bauthor{\bsnm{{Ade}}, \binits{P.A.R.}},
\bauthor{\bparticle{et} \bsnm{al.}}:
\bjtitle{\aap}
\bvolume{571},
\bfpage{16}
(\byear{2014})
\end{barticle}
\endbibitem

\bibitem[\protect\citeauthoryear{{Rauch}}{1998}]{Rauch1998}
\begin{barticle}
\bauthor{\bsnm{{Rauch}}, \binits{M.}}:
\bjtitle{\araa}
\bvolume{36},
\bfpage{267}
(\byear{1998})
\end{barticle}
\endbibitem

\bibitem[\protect\citeauthoryear{{Sluse} et~al.}{2011}]{Sluse2011}
\begin{barticle}
\bauthor{\bsnm{{Sluse}}, \binits{D.}},
\bauthor{\bsnm{{Schmidt}}, \binits{R.}},
\bauthor{\bsnm{{Courbin}}, \binits{F.}},
\bauthor{\bsnm{{Hutsem{\'e}kers}}, \binits{D.}},
\bauthor{\bsnm{{Meylan}}, \binits{G.}},
\bauthor{\bsnm{{Eigenbrod}}, \binits{A.}},
\bauthor{\bsnm{{Anguita}}, \binits{T.}},
\bauthor{\bsnm{{Agol}}, \binits{E.}},
\bauthor{\bsnm{{Wambsganss}}, \binits{J.}}:
\bjtitle{\aap}
\bvolume{528},
\bfpage{100}
(\byear{2011})
\end{barticle}
\endbibitem

\bibitem[\protect\citeauthoryear{{Snow} and {McCall}}{2006}]{Snow2006}
\begin{barticle}
\bauthor{\bsnm{{Snow}}, \binits{T.P.}},
\bauthor{\bsnm{{McCall}}, \binits{B.J.}}:
\bjtitle{\araa}
\bvolume{44},
\bfpage{367}
(\byear{2006})
\end{barticle}
\endbibitem

\bibitem[\protect\citeauthoryear{{Zeldovich} and {Novikov}}{}]{Zeldovich1983}
\begin{botherref}
\oauthor{\bsnm{{Zeldovich}}, \binits{I.B.}},
\oauthor{\bsnm{{Novikov}}, \binits{I.D.}}:
{Relativistic Astrophysics. Volume 2 - The Structure and Evolution of the
  Universe. Chicago, Il, University of Chicago Press, (1983)}
\end{botherref}
\endbibitem

\end{thebibliography}

\section*{Appendix}
In Appendix we obtain an analytic approximations
of $g(\kappa, \delta)$  determined 
in Eqs.~(\ref{g_k_d}) and (\ref{s_r_d})
and represented as a set of sections continuously sewed
together in surroundings of definite boundary lines
$\kappa (\delta)$ (see below) at fixed $z_c$ and $z_q$.
Such an approach allows us to derive an appropriate set 
of simple reciprocal  functions $\kappa (g, \delta)$   
and to use them in the integrations 
of Eqs.~(\ref{P(g)}) and (\ref{P(f)}). 

Firstly,  using Eq.~(\ref{rho_kappa})
one can define 
two basic boundary functions on $\delta$ 
at the plane $\delta - \kappa$:    
\begin{align}
&\kappa_1 = \frac{D(z_c)}{D(z_q)}|\delta - 1|, &  
			&\kappa_2 = \frac{D(z_c)}{D(z_q)}(\delta + 1). 
\label{k1_2}
\end{align}
These functions  
correspond to the boundaries $\rho=|\delta -1|$ and
$\rho=\delta + 1$, respectively, 
at the plane $\delta - \rho$, which
separate the  {\it crescent-like coverage} region (iii) 
from the regions (ii), (iv) and (i)
(see Fig.~\ref{types}). 

Then, 
to obtain simple approximations within the region (iii),
we introduce two additional interfacing 
dependences on $\delta$ 
\begin{align}
&\kappa_{1/3} = \frac{2}{3}\kappa_1 + \frac{1}{3}\kappa_2 &
			&\kappa_{2/3} = \frac{1}{3}\kappa_1 + \frac{2}{3}\kappa_2, 
\label{k1/3_2/3}
\end{align}
and using Eqs.~(\ref{s_r_d}) and (\ref{rho_kappa})
calculate corresponding accurate functions    
\begin{align}
&g_{1/3} = g(\kappa_{1/3},\delta) &
			&g_{2/3} = g(\kappa_{2/3},\delta).
\label{g1/3_2/3}
\end{align}
Note, that all defined quantities depend on 
$z_q$ and $z_c$, which are incorporated in the dependences 
$\kappa (\delta)$.  

To obtain the domain-like approximations for 
$g(\kappa, \delta)$ 
it is convenient to introduce two basic zones: 
(a) with $\delta \geq 1$
and (b) with $\delta < 1$.  

In the zone (a) at $\delta \geq 1$ we have:
\begin{multline}
g(\kappa, \delta) = \\
\left\{
\begin{array}{ll}
0; & 0 < \kappa < \kappa_1  \\
g_{1/3}\left( {\kappa - \kappa_1 \over \kappa_{1/3}-\kappa_1} \right)^{3/2}; & 
\kappa_1 \leq \kappa < \kappa_{1/3} \\
\, & \, \\
\lefteqn{g_{1/3} +} & \\
(g_{2/3}-g_{1/3}) {\kappa - \kappa_{1/3} \over \kappa_{2/3}-\kappa_{1/3}} ; & 
\kappa_{1/3} \leq \kappa < \kappa_{2/3} \\
\, & \, \\
1-(1-g_{2/3})\left( {\kappa_2 - \kappa \over
\kappa_2 - \kappa_{2/3}} \right)^{3/2}; & 
\kappa_{2/3} \leq \kappa < \kappa_2 \\
\, & \, \\
1; & \kappa_2 \leq \kappa,
\end{array}
\right.
\label{g_a}
\end{multline}
where  five sections consequently  correspond (from top to bottom):
1-st section to the region (iv) at $\rho < \delta -1$; 
2-nd, 3-d and 4-th sections to (iii)
at $\delta -1 \leq \rho < \delta + 1$; 5-th section to (i)
at $\delta + 1 \leq \rho$.

In the zone (b) at $\delta < 1$ we have:
\begin{multline}
g(\kappa,\delta) = \\
\left\{
\begin{array}{ll}
\rho^2(\kappa); & 0 < \kappa \leq \kappa_1 \\
\, & \, \\
\lefteqn{{g_{1/3}- \Delta \over \kappa_{1/3}^{3/2} - \kappa_1^{3/2}}\, 
\kappa^{3/2} +
{\kappa_{1/3}^{3/2} \Delta - \kappa_1^{3/2} g_{1/3} 
\over \kappa_{1/3}^{3/2}-\kappa_1^{3/2}};}  & \, \\ 
\, & \kappa_1 \leq \kappa < \kappa_{1/3} \\
\, & \, \\
g_{1/3}+(g_{2/3}-g_{1/3}){\kappa - \kappa_{1/3} \over \kappa_{2/3}-\kappa_{1/3}}; & 
\kappa_{1/3} \leq \kappa < \kappa_{2/3} \\
\, & \, \\
1-(1-g_{2/3})\left( {\kappa_2 - \kappa \over \kappa_2 - \kappa_{2/3}} 
\right)^{3/2}; & \kappa_{2/3} \leq \kappa < \kappa_2 \\
\, & \, \\
1; & \kappa_2 \leq \kappa,
\end{array}
\right.
\label{g_b}
\end{multline}
where $\Delta=(1-\delta)^2$,  five 
sections consequently  correspond:
1-st upper section to the region (ii) at $\rho < 1-\delta$; 
2-nd, 3-d and 4-th sections to (iii)
at $1- \delta \leq \rho < 1+ \delta$; 5-th section to (i)
at $1+ \delta  \leq \rho$.
Note that a difference between the exact coverage 
factor  $g(\kappa,\delta)$   and the approximations 
given by Eqs.~(\ref{g_a}) and (\ref{g_b})
does not exceed $0.02$. 

Finally, we can obtain an approximation 
for the reciprocal function $\kappa(g,\delta)$,
treating the same two zones: 
(a) with $\delta \geq 1$
and (b) with $\delta < 1$.  

Thus in the zone (a) at $\delta \geq 1$ we have:
\begin{multline}
\kappa(g,\delta) = \\
\left\{
\begin{array}{ll}
\kappa_1 +(\kappa_{1/3} - \kappa_1)\left( {g \over g_{1/3}} \right)^{2/3}; &
 0 \leq g < g_{1/3} \\
\, & \, \\
\kappa_{1/3} + (\kappa_{2/3} -\kappa_{1/3}){g - g_{1/3} \over g_{2/3} - g_{1/3}}; & 
g_{1/3} \leq g < g_{2/3} \\
\, & \, \\
\kappa_2-(\kappa_2 - \kappa_{2/3})\left({1-g \over 1-g_{2/3}} \right)^{2/3}; & g_{2/3} \leq g \leq 1,
\end{array}
\right.
\label{k_a}
\end{multline}
where all three sections 
correspond to the region (iii) and its boundaries with 
the regions (iv) at $g=0$  and (i) at $g=1$.

In the zone (b) at $\delta < 1$ employing Eq.~(\ref{rho_kappa}) 
we have:
\begin{multline}
\kappa(g,\delta) = \\
\left\{
\begin{array}{ll}
{D(z_c) \over D(z_q)}\, \sqrt{g}; & 0 \leq g < \Delta \\
\, & \, \\
\lefteqn{ \left[
{ \kappa_{1/3}^{3/2}\, (g - \Delta) + \kappa_1^{3/2}\, (g_{1/3}-g) 
\over g_{1/3} - \Delta}  \right]^{2/3};} & \, \\   
\, &  \Delta \leq g < g_{1/3}  \\
\, & \, \\
\kappa_{1/3} + (\kappa_{2/3} -\kappa_{1/3}){g - g_{1/3} \over g_{2/3} - g_{1/3}}; & 
g_{1/3} \leq g < g_{2/3} \\
\, & \, \\
\kappa_2-(\kappa_2 - \kappa_{2/3})\left({1-g \over 1-g_{2/3}} \right)^{2/3}; & g_{2/3} \leq g \leq 1,
\end{array}
\right.
\label{k_b}
\end{multline}
where  as in Eq.~(\ref{g_b}) 
$\Delta=(1-\delta)^2$, 
and the first upper section  
corresponds to the region (ii), 
all three others to  (iii).

\end{document}